\documentclass[
aps,
prl,
reprint,
superscriptaddress,
floatfix,
nofootinbib,
twocolumn,
longbibliography
]{revtex4-2}

\usepackage[dvipdfmx]{graphicx}

\usepackage{amsmath,amssymb,amsfonts,physics,graphicx,float}
\usepackage{amsmath}
\usepackage{stmaryrd}
\graphicspath{{pic/}}
\usepackage[setpagesize=false]{hyperref}
\usepackage{cancel}

\allowdisplaybreaks[4]

\usepackage{ulem}
\usepackage{color}
\definecolor{ar}{rgb}{1.0, 0.01, 0.24}
\definecolor{al}{rgb}{0.82, 0.1, 0.26}
\definecolor{ev}{rgb}{0.56, 0.0, 1.0}
\def\be{\begin{eqnarray}}\def\ee{\end{eqnarray}}

\allowdisplaybreaks

\begin{document}

\author{Mamiya Kawaguchi}
\email{mamiya@aust.edu.cn}
\affiliation{
Center for Fundamental Physics, School of Mechanics and Photoelectric Physics,
Anhui University of Science and Technology, Huainan, 232001, China}

\author{Masayasu Harada}
\email{harada@hken.phys.nagoya-u.ac.jp}
\affiliation{Kobayashi-Maskawa Institute for the Origin of Particles and the Universe, Nagoya University, Nagoya, 464-8602, Japan}
\affiliation{Department of Physics, Nagoya University, Nagoya 464-8602, Japan}
\affiliation{Advanced Science Research Center, Japan Atomic Energy Agency, Tokai 319-1195, Japan}

\author{Yong-Liang Ma}
\email{ylma@nju.edu.cn}
\affiliation{School of Frontier Sciences, Nanjing University, Suzhou 215163, China}

\title{
Origin of hadron mass from gravitational $D$-form factor and neutron star measurements
}

\begin{abstract}
Clarifying the origin of hadron mass is one of the fundamental problems in particle physics, relevant from hadronic scales to astrophysical observations. At low energies, this issue is reflected in the decomposition of the hadron mass into chiral-variant and -invariant components. In this letter, we propose a method to extract the chiral invariant mass from the gravitational $D$-form factor under the assumption of the lightest-sigma meson dominance. Focusing on the nucleon, we show that a sizable chiral invariant mass is required to reproduce lattice QCD data, consistent with neutron star constraints.

\end{abstract}

\pacs{}

\maketitle

{\it Introduction}---Understanding the origin of the hadron mass is one of the central open questions in particle and nuclear physics~\cite{Wilczek:2006eg,Wilczek:2012sb}.  
At the low-energy regions, the spontaneous chiral symmetry breaking is conventionally recognized as the dominant mechanism generating hadron masses~\cite{Nambu:1961tp,Nambu:1961fr}  and the nucleon mass arises entirely from the chiral non-invariant quark condensate.  
However, lattice QCD simulations~\cite{Aarts:2015mma,Aarts:2017rrl,Aarts:2018glk} and QCD sum rule analyses~\cite{Kim:2020zae,Kim:2021xyp,Lee:2023ofg} indicate that the nucleon mass also contains a chiral invariant contribution that is independent of the quark condensate.
Thus, the nucleon mass can be decomposed into a chiral variant part and a chiral invariant part~\cite{Detar:1988kn}. 
This mass decomposition based on chiral transformation is not limited to the nucleon but also applies to other hadrons.
Clarifying this mass decomposition is essential for understanding the fundamental origin of hadron masses.

This fundamental problem of the origin of hadron mass is further connected to astrophysics, particularly the physics of neutron stars (NSs). In nuclear matter, the Yukawa interaction between the nucleon and the scalar meson provides the attractive force, which strongly influences the stiffness of the equation of state (EOS)~\cite{Yamazaki:2019tuo,Minamikawa:2020jfj}. Importantly, the strength of this Yukawa coupling is affected by the chiral invariant mass of the nucleon (see Eq.~\eqref{yukawa}).
In an extremely high dense system, the chiral symmetry is expected to gradually restore, and the mass arising from spontaneous chiral symmetry breaking is reduced, while the chiral invariant mass remains~\cite{Aarts:2015mma,Aarts:2017rrl,Aarts:2018glk}, thereby modifying the strength of the Yukawa interaction. 
Consequently, this density dependence of the nucleon mass is reflected in the NS EOS.
Recent studies based on the parity-doublet nucleon model, in which the chiral invariant mass $m_0$ is incorporated~\cite{Detar:1988kn,Jido:2001nt}, suggest that, satisfying observational constraints on NS masses and radii requires a substantial portion of the nucleon mass to originate from the chiral invariant component, typically indicating $m_0=500$-$900$ MeV~\cite{Minamikawa:2020jfj,Minamikawa:2023eky, Gao:2024lzu}. 
Although the origin of $m_0$ is not specified---potentially stemming from scale symmetry violation triggered by the gluon condensate~\cite{Sasaki:2011ff,Ma:2018xjw}---its value is constrained by NS measurements.

In recent years, the internal structures of the nucleon have attracted significant attention, motivated by experimental studies of the stress distribution inside the proton~\cite{Burkert:2018bqq,Burkert:2021ith,Duran:2022xag}. These internal properties are connected with the gravitational form factors (GFFs), which are defined through the matrix element of the energy-momentum tensor $\Theta^{\mu\nu}$ as
\be
&& \langle N(p',s')| \Theta^{\mu\nu} | N(p,s)  \rangle \nonumber\\
& &
=
\bar u^{s'}(p') 
\Bigg[
A_{N}(t)
\frac{\bar P^\mu \bar P^\nu}{m_{N}}
+J_{N}(t)
\frac{
i\sigma^{\mu\rho}q_\rho
\bar P^\nu
+
i\sigma^{\nu\rho}q_\rho
\bar P^\mu}{2 m_{N}}
\nonumber\\
&& \qquad\qquad\;\;\; 
+
D_{N}(t)
\frac{(q^\mu q^\nu-g^{\mu\nu}q^2    )}{4 m_{N}}
\Bigg]
u^s(p)
e^{-iq\cdot x},
\ee
where $\sigma_{\mu\nu}=\frac{i}{2}[\gamma^\mu,\gamma^\nu]$; $\bar P^\mu=(p^\mu+p'^\mu)/2$ is the average momentum, $q^\mu=p'^\mu-p^\mu$ is the momentum transfer, and $t=-q^2$; $m_N$ is a nucleon mass; $u(p,s)$ is the Dirac spinor with spin $s$. $A_N(t)$ and $J_N(t)$ are related to the total mass and spin distributions, respectively, whereas the $D_N(t)$ is directly linked to the internal forces and mechanical properties of the nucleon. 
The GFFs of other hadrons can also be formulated in a similar manner. 
Despite the relevance of the $D$-form factor to the internal structure of hadrons, it remains poorly understood. Although lattice QCD simulations~\cite{Shanahan:2018pib,Shanahan:2018nnv,Pefkou:2021fni,Hackett:2023nkr,Hackett:2023rif,Wang:2024lrm} and various recent theoretical approaches~\cite{Yao:2024ixu,Cao:2024zlf,Broniowski:2025ctl,Xing:2025uwn} have made progresses in clarifying it (see also reviews~\cite{Polyakov:2018zvc,Burkert:2023wzr}), how $D(t)$ connects to fundamental hadron properties, such as the origin of the hadron mass, remains unclear.

In this work, we aim to extract the chiral invariant mass of a hadron from its $D$-form factor. Previous studies have shown that the meson dominance is realized in the GFFs of the pion and the nucleon at low energies~\cite{Broniowski:2024oyk,Broniowski:2025ctl,Stegeman:2025sca}. Motivated by this observation, we adopt the lightest scalar meson dominance picture throughout our analysis. Identifying this lightest scalar meson with the sigma meson associated with spontaneous chiral symmetry breaking, we argue that the $D$-form factor of a hadron is, in general, sensitive to the chiral invariant mass of the corresponding hadron. As a concrete demonstration of this idea, we apply our analysis to the nucleon.
We first employ a conventional chiral effective model containing only the positive-parity nucleon, whose mass arises solely from spontaneous chiral symmetry breaking, and compare the resulting $D$-form factor with the recent lattice QCD simulation. We then extend this framework to the parity doublet nucleon structure in order to incorporate the chiral invariant mass, and examine its impact on the $D$-form factor.
\\

{\it $D$-form factor and chiral symmetry
}---Recent analyses of the GFFs of the pion and the nucleon suggest that the meson dominance is realized in the low-energy regime~\cite{Broniowski:2024oyk,Broniowski:2025ctl,Stegeman:2025sca}. Given this fact, one generally expects the GFFs of hadrons, especially the $D$-form factor, to be governed by the exchange of the lightest scalar meson when the energy region is restricted to below $1\,{\rm GeV}$. In this scenario, the momentum-transfer dependence of the $D$-form factor for a hadron takes the generic form
\begin{equation}
D_{H }(t)\sim g_{HH\sigma} \frac{1}{m_\sigma^2+t}
\end{equation}
where $m_\sigma$ is the mass of the lightest scalar meson and $g_{HH\sigma}$ denotes the coupling constant for the three-point interaction between the hadron and the scalar meson.

Supposing that this scalar meson is identified as the sigma meson associated with the chiral transformation, the coupling strength of $g_{HH\sigma}$ is constrained by the chiral symmetry.
Specifically, the coupling constant is related to the hadron mass as
\begin{equation}
g_{HH\sigma} \propto m^{(H)} - m^{(H)}_0,
\label{gene_coupling}
\end{equation}
where $m^{(H)}$ is the entire hadron mass and $m_0^{(H)}$ represents its chiral invariant component. 
Note that this expression is meant only as an illustrative form, and the actual situation makes this form more complicated.
The key point is that, once chiral symmetry is restored, the two masses coincide, $m^{(H)} = m_0^{(H)}$, which results the coupling to zero. 
The $D$-form factor is therefore sensitive to the chiral invariant mass of the corresponding hadron. In this picture, a sizable chiral invariant mass reduces the magnitude of the $D$-form factor.

In this letter, focusing on the nucleon, we demonstrate how the chiral invariant mass can be extracted from the nucleon $D$-form factor.
\\

{\it A linear sigma model with a positive-parity nucleon}---As a concrete demonstration of the discussion in the previous section, we begin with a conventional linear sigma model describing a positive-parity nucleon coupled to the lightest mesons, in which the chiral invariant mass is absent,
\begin{equation}
{\cal L} = {\cal L}_{N} + {\cal L}_M.
\label{Lag LSM}
\end{equation}
The nucleon and meson sectors are given by
\be
{\cal L}_{N} & = & \bar N_r i\gamma^\mu \partial_\mu N_r
+
\bar N_l i\gamma^\mu \partial_\mu N_l
\nonumber\\
& & {}
-g\left(
\bar N_r M^\dagger N_l +
\bar N_l M N_r\right), \nonumber\\
{\cal L}_M & = &
\frac{1}{4} \, \mbox{tr} \left[ \partial^\mu M^\dag \, \partial_\mu M \right]  
-V\left( M\right),
\ee
where $N=(p,n)$ denotes the positive parity nucleon and $M=\sigma+i\sum_{a=1}^{3}\pi^a \tau^a$ denotes the chiral meson field multiplet in the linear representation. The mesonic potential $V(M)$ induces the spontaneous chiral symmetry breaking. 

Under the chiral symmetry  $SU(2)_L\times SU(2)_R$, the right (left) handed nucleon field and the meson field transform  as
\begin{equation}
\begin{split}
N_{r(l)} \to g_{R(L)} N_{r(l)} ,\quad
M  \to g_L M g_R^\dagger,
\end{split}
\end{equation}
where $g_{L,R} \in SU(2)_{L,R}$. In the present analysis, we neglect the explicit chiral symmetry breaking due to the current quark masses. Then the Lagrangian~(\ref{Lag LSM}) is invariant under the chiral transformation.
When the chiral symmetry is spontaneously broken by the vacuum expectation value of the sigma meson field $\langle0|\sigma |0 \rangle =f_\pi$, the nucleon acquires a finite mass through the Yukawa interaction with coupling $g$,
\begin{align}
    m_N= g f_\pi.
\label{nucelon_mass}
\end{align}
In the linear sigma model with only the positive-parity nucleon, the nucleon mass originates entirely from the chiral symmetry breaking. As a result, the Yukawa coupling is determined solely by the nucleon mass.

In this framework, the energy–momentum tensor (EMT) takes
the form
\be
\Theta^{\mu\nu} & = & 
\frac{1}{2}{\bar N_{r}}i(\gamma^\mu{ \partial^\nu}
+\gamma^\nu{ \partial^\mu}
) N_{r}
+\frac{1}{2}{\bar N_{l}}i(\gamma^\mu{\partial^\nu}
+
\gamma^\nu{\partial^\mu}
)
N_{l}
\nonumber\\
& &{} 
+\frac{1}{2} \, \mbox{tr} \left[ \partial^\mu M^\dag \, \partial^\nu  M \right]
-g^{\mu\nu} {\cal L}
\nonumber\\
& &{} 
- \frac{1}{12}(\partial^\mu \partial^\nu -g^{\mu\nu}\partial^2 ){\rm tr}[M M^\dagger].
\ee
This EMT is connected to the trace anomaly via $\Theta^\mu_{\;\mu} =\partial_\mu j^\mu_D$ with the dilatation current $j^\mu_D$, and it couples directly to the sigma meson in the spontaneous chiral symmetry broken phase, 
\begin{equation}
\begin{split}
\langle 0|\Theta^{\mu\nu}| \sigma(p) \rangle & =
\frac{f_\pi}{3}(p^\mu p^\nu -g^{\mu\nu} m_{\sigma}^2)e^{-ip\cdot x}.
\end{split}
\label{matrix_element}
\end{equation}
This matrix element combined with the Yukawa interaction gives rise the nucleon $D$-form factor.
\\

\begin{figure}[h]
\includegraphics[scale=0.35]{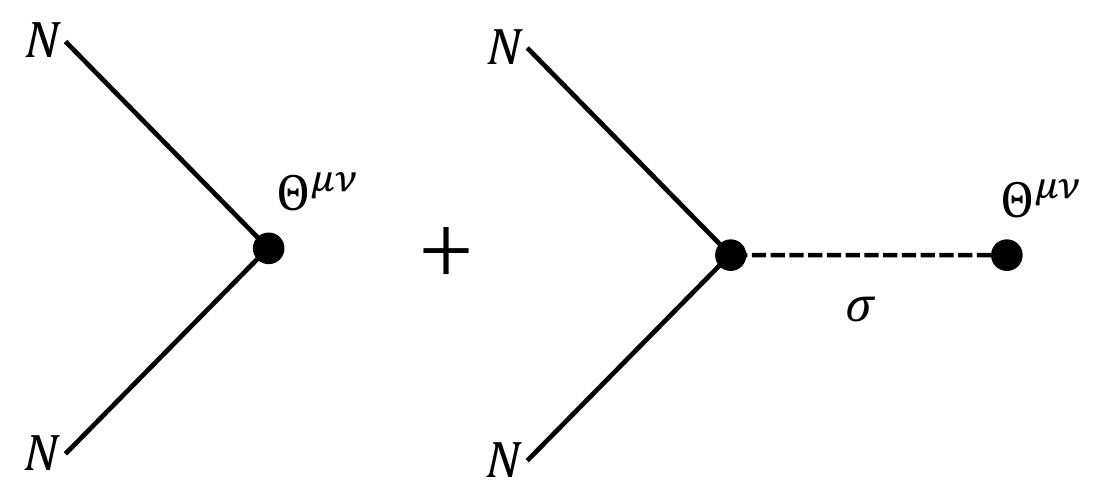}
\caption{Feynman diagrams for the nucleon GFFs.
  The first contribution arises from the contact interaction $NN\Theta^{\mu\nu}$ and the second is derived by the Yukawa interaction $NN\sigma$ mediated by the lightest-sigma meson. 
  }
  \label{diagram}
\end{figure}

{\it$D$-from factor in the linear sigma model}---Under the lightest–sigma meson dominance, the nucleon GFFs are straightforwardly evaluated through the diagrams shown in Fig.~\ref{diagram}.
The contact term corresponding to the first diagram in Fig.~\ref{diagram} yields $A(t)=1,\quad J(t)=1/2$.
The second diagram arises from the lightest sigma-meson exchange through the Yukawa interaction together with the nonzero matrix element in Eq.~\eqref{matrix_element}, generating the $D$-form factor,
\be
D_N^{(\text{LSM})}(t) & = &{} -\frac{4 m_{N}^2}{3} 
\frac{
1
}{m_{\sigma}^2+t}.
\label{Dt_conv_N}
\ee
In obtaining this expression, we have used the mass formula related to the Yukawa coupling in Eq.~\eqref{nucelon_mass}. Consequently, the $D$-form factor is expressed in terms of the entire nucleon mass and the lightest sigma-meson mass. We denote this result by $D_N^{(\text{LSM})}(t)$. 
Note that a similar expression also appears in other effective model descriptions based on the dilaton effective theory~\cite{Stegeman:2025sca}.

\begin{figure}[h]
\includegraphics[scale=0.35]{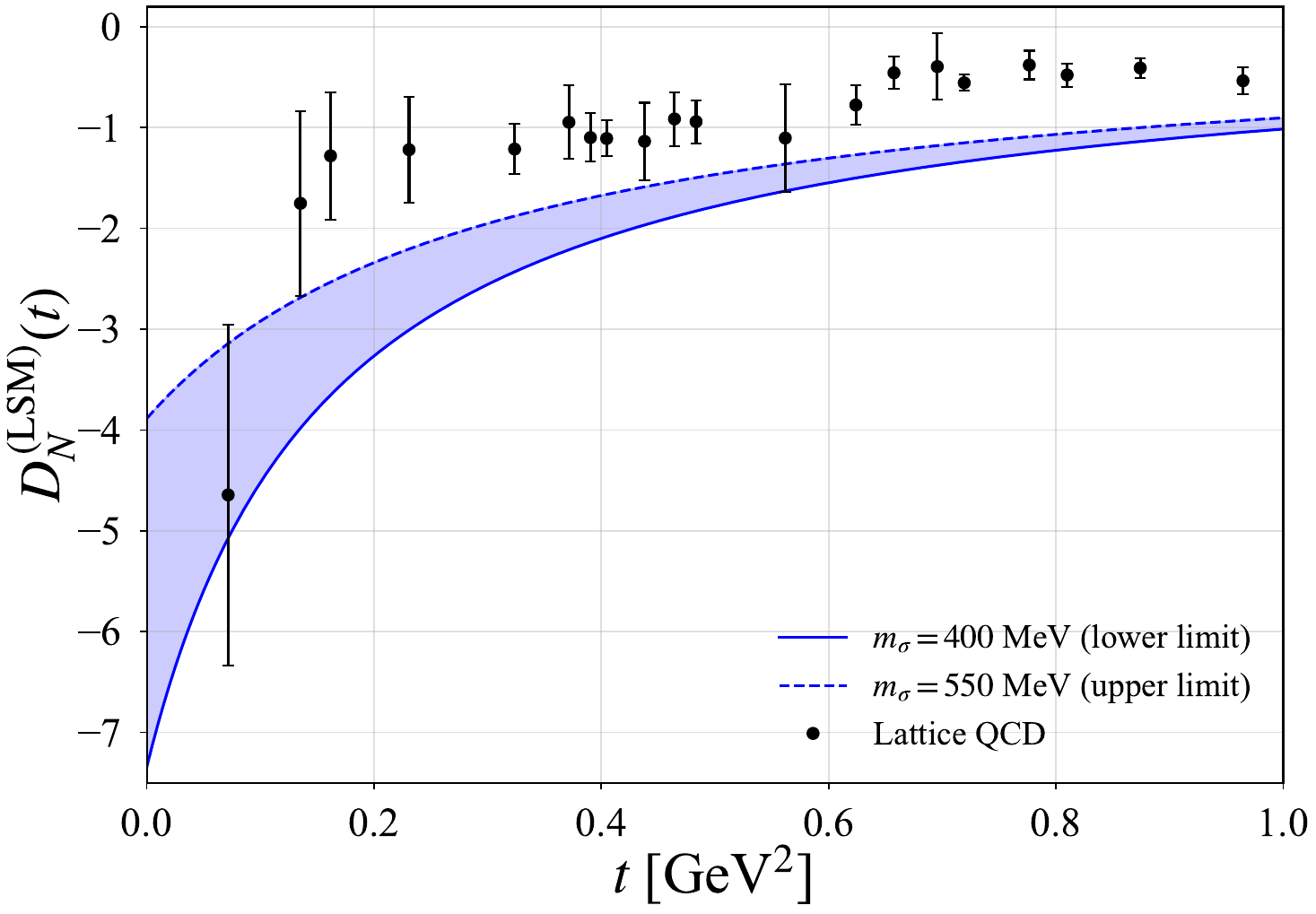}
\caption{Momentum transfer dependence of $D$-form factor in the chiral framework with only the positive-parity nucleon, compared with the lattice QCD data~\cite{Hackett:2023rif}.
}
  \label{Dform_LSM}
\end{figure}

With the analytic expression~\eqref{Dt_conv_N} at hand, we now perform a numerical evaluation. The nucleon mass is taken to be $m_N=939\,{\rm MeV}$. Since the mass of the lightest scalar-meson identified with $f_0(500)$ is not well determined, we consider the range $m_\sigma = 400\text{–}550\;{\rm MeV}$, corresponding to the values listed in the Particle Data Group~\cite{ParticleDataGroup:2024cfk}.
With these physical values, we plot the momentum-transfer dependence of $D_N^{(\text{LSM})}(t)$ in Fig.~\ref{Dform_LSM} and compare it with the 
lattice QCD data for the proton $D$-form factor obtained at $m_\pi=170\,{\rm MeV}$, near the physical point~\cite{Hackett:2023rif}. 
As shown in this figure, the $D$-form factor without the chiral invariant mass deviates from the lattice data and fails to provide satisfactory agreement.

In addition, in the forward limit, our result yields $D_N^{(\text{LSM})}(0) =-3.89\text{ to} -7.35 $. The magnitude of this estimate is, however, somewhat larger than a variety of theoretical studies, most of which yield values close to $D_N(0)\sim -3$~\cite{Won:2022cyy,Yao:2024ixu,Cao:2024zlf,Broniowski:2025ctl,Xing:2025uwn}. Specifically, in numerical analyses, the continuum Schwinger method reports $D_N(0)=-3.11(1)$~\cite{Yao:2024ixu} while the dispersion-relation analysis yields $D_N(0)=-3.38^{+0.34}_{-0.35}$~\cite{Cao:2024zlf}.

These deviations imply that some ingredients are missing, such as a sizable chiral invariant mass---another mechanism of mass generation other than chiral symmetry breaking---as suggested in the following.
\\

{\it Chiral invariant mass mechanism of nucleon}---Following the parity doublet concept~\cite{Detar:1988kn,Jido:2001nt}, we incorporate the chiral invariant mass into the linear sigma model in Eq.~\eqref{Lag LSM}. 
We then extend the nucleon sector by including both the positive-parity nucleon ($\psi_{1}$) and the negative-parity nucleon ($\psi_{2}$), 
\be
{\mathcal L}_{N}^{\rm (PDM)} & = & 
{\bar\psi_{1r}}i\gamma^\mu{ \partial_\mu}\psi_{1r}
+{\bar\psi_{1l}}i\gamma^\mu{\partial_\mu}\psi_{1l} \nonumber\\
& &{} +{\bar\psi_{2r}}i\gamma^\mu{\partial_\mu}\psi_{2r}
+{\bar\psi_{2l}}i\gamma^\mu{\partial_\mu}\psi_{2l} \nonumber\\
& &{} - {m}_0\, \left[{\bar\psi_{1l}}\psi_{2r}-{\bar\psi_{1r}}\psi_{2l}
-{\bar\psi_{2l}}\psi_{1r}+{\bar\psi_{2r}}\psi_{1l}\right]
\nonumber\\
& &{} 
-g_1\left[{\bar\psi_{1r}} M^\dag \psi_{1l}+{\bar\psi_{1l}} M\psi_{1r}\right] \nonumber\\
& &{} -g_2 \left[{\bar\psi_{2r}} M \psi_{2l} +{\bar\psi_{2l}} M^\dag \psi_{2r}\right],
\ee
where $g_{1,2}$ denote the Yukawa couplings in the parity doublet framework. 
Under the chiral transformation, the nucleon fields transform as
\be
& & \psi_{1r} \to g_R \psi_{1r}, \quad
\psi_{1l} \to g_L  \psi_{1l}, \nonumber\\
& & \psi_{2r} \to g_L \psi_{2r}, \quad
\psi_{2l} \to g_R  \psi_{2l},
\ee
which is known as the mirror assignment. Owing to the parity-doublet structure, a chiral invariant nucleon mass $m_0$ is allowed in the Lagrangian. The presence of the chiral invariant mass induces the mixing between $\psi_1$ and $\psi_2$. After diagonalization, the physical nucleon masses become 
\begin{equation}
\begin{split}
m_{\pm} &=\frac{1}{2}\left(
\pm(g_1 - g_2) f_\pi + \sqrt{ f_\pi^2( g_1+g_2 )^2 + 4  m_0^2 
}
\right).
\end{split}
\end{equation}
Here, $m_+=939\,{\rm MeV}$ corresponds to the positive parity nucleon $N(939)$, and $m_-=1535\,{\rm MeV}$ corresponds to the negative parity nucleon $N^*(1535)$.
In contrast to the nucleon mass~\eqref{nucelon_mass} which arises totally from chiral symmetry breaking, the parity-doubled nucleon mass is described by two distinct contributions: one from spontaneous chiral symmetry breaking tagged with $f_\pi$, and the other from the chiral invariant mass $m_0$.

The Yukawa coupling between the physical positive-parity nucleon and the sigma meson is given by
\be
g_{NN\sigma } 
& = &
\frac{1}{f_\pi}
\left(
m_{+}-\frac{2m_0^2}{m_{+} + m_{-}}
\right),
\label{yukawa}
\ee
indicating that the chiral invariant mass explicitly enters the Yukawa coupling and reduces its strength, as discussed around Eq.~\eqref{gene_coupling}.

It is interesting to note that the Yukawa interaction plays an essential role in NS physics. This coupling provides the attractive force in nuclear matter, and strongly influences the stiffness of the nuclear EOS~\cite{Yamazaki:2019tuo,Minamikawa:2020jfj}. The recent analyses based on the parity doublet model suggest that the chiral invariant mass accounts for most of the nucleon mass, typically lying in the range $m_0=500\text{-}900\text{ MeV}$~\cite{Minamikawa:2020jfj,Minamikawa:2023eky, Gao:2024lzu}.
\\

{\it Reduction of $D$-form factor by the chiral invariant mass}---Based on the parity doublet description, the $D$-form factor is now expressed as
\be
D_N^{\rm (PDM)}(t) & = &{} -\frac{4 m_{+}}{3} 
\frac{
g_{NN\sigma} f_\pi   
}{m_{\sigma}^2+t} \nonumber\\
& = &
\left( 1 - \frac{ 2 m_0^2 }{ m_+ \left( m_+ + m_- \right) } \right) D_N^{\rm(LSM)}(t),
\ee
where $D_N^{(\text{LSM})}(t)$ is identical to the result obtained in Eq.~\eqref{Dt_conv_N}.
This clearly shows that the chiral invariant mass reduces the overall magnitude of the $D$-form factor. 

\begin{figure}[h]
\includegraphics[scale=0.35]{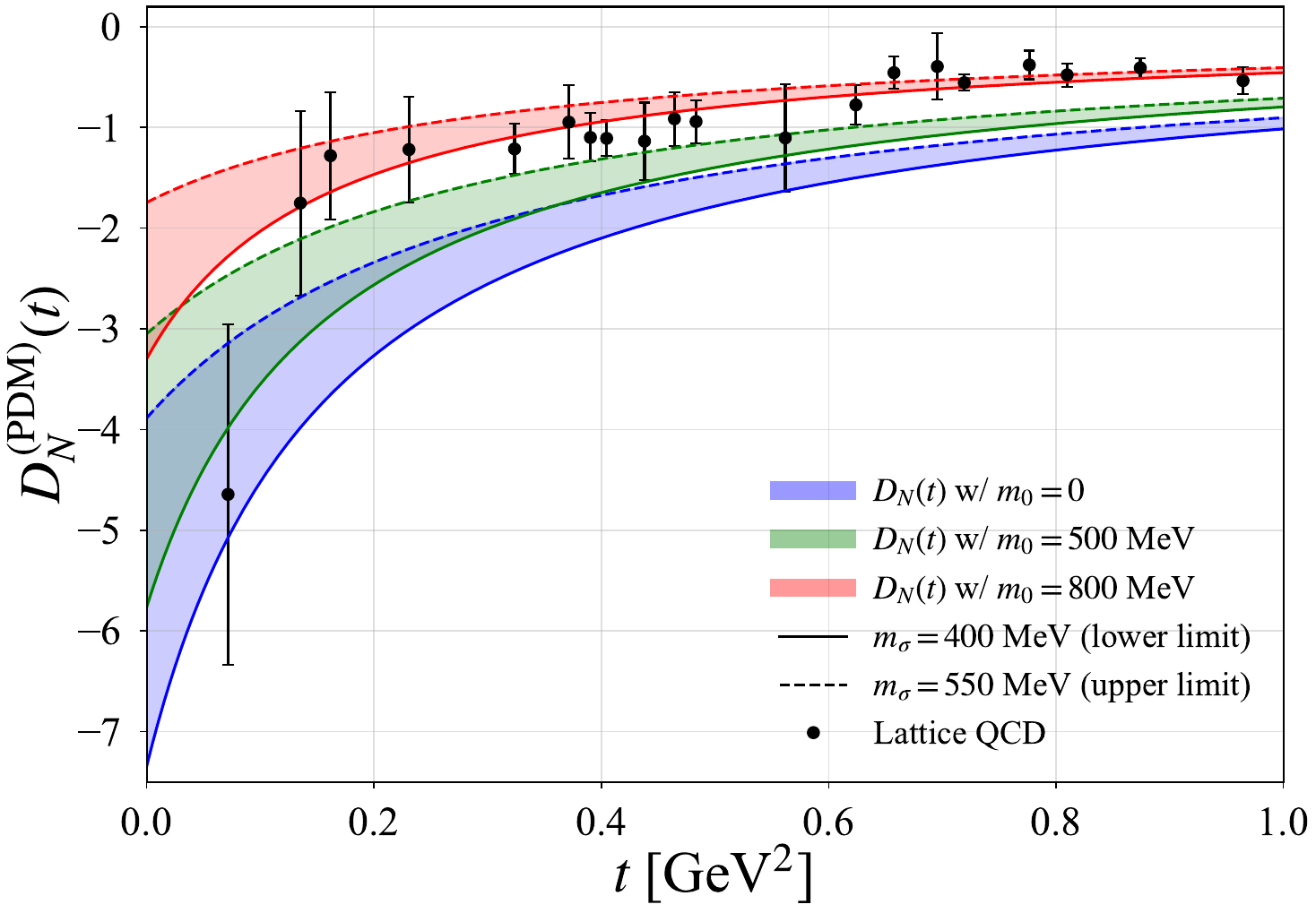}
  \caption{Reduction of the nucleon $D$-form factor due to the chiral invariant mass, compared with the lattice QCD data~\cite{Hackett:2023rif}.
  }
  \label{CIM_Dt}
\end{figure}

To quantify the reduction induced by the chiral invariant mass $m_0$, we show the numerical behavior of $D_N^{\rm (PDM)}(t)$ in Fig.~\ref{CIM_Dt}. Following NS analyses based on the parity doublet model, we consider two representative values, $m_0=500\,{\rm MeV}$ and $m_0=800\,{\rm MeV}$, which lie within the commonly discussed range for the chiral invariant mass.  Note that for $m_0=0$, the parity doublet result coincides with that of the linear sigma model with only the positive parity nucleon shown in Fig.~\ref{Dform_LSM}.  
Figure~\ref{CIM_Dt} clearly shows that once a finite chiral invariant mass is introduced, the chiral invariant mass reduces the magnitude of $D(t)$, causing the parity doublet result to move closer to the lattice data.
In particular, for $m_0=800{\rm MeV}$, the result closely follows the lattice data, especially in the large momentum-transfer region where the uncertainty of the scalar-meson mass becomes less significant. Furthermore, the forward limit becomes $D_N(0)=-1.75\text{ to}-3.30$, which is in much closer agreement with other theoretical estimates than the result obtained in the absence of the chiral invariant mass.\\

{\it Conclusions}---
In this letter, we have pointed out that, under the sigma meson dominance, the $D$-form factor of a hadron is directly linked to the chiral invariant mass of the corresponding hadron.
As a demonstration, we have focused on the nucleon and examined its $D$-form factor within the parity doublet model, comparing the result with lattice QCD data.
Our analysis clearly shows that the chiral invariant mass of the nucleon reduces the magnitude of the $D$-form factor. This reduction stems from the $m_0$ dependence of the Yukawa coupling.
This sensitivity suggests that the $D$-form factor serves as a novel probe of the nucleon mass structure.
When the chiral invariant mass takes a relatively large 
value, such as $m_0 = 800,\mathrm{MeV}$, the resulting $D$-form factor becomes consistent with current lattice QCD data, except for the lowest energy point.
 Remarkably, the resulting value of $m_0$ also aligns with recent NS studies based on the parity doublet framework. 

We note that our analysis has been restricted to the low energy region below $1\,\mathrm{GeV}$, where we assumed the lightest-scalar meson dominance.
At higher energies, additional meson exchanges can contribute to the $D$-form factor. 
In particular, heavier isosinglet scalar mesons such as $f_0(980)$, $f_0(1370)$, $f_0(1500)$, and $f_0(1710)$, as well as tensor mesons including $f_2(1270)$, are expected to contribute to the GFFs. Nevertheless, the dominance of the lightest scalar meson governs the leading contribution in the low energy region and does not alter our qualitative conclusion, since heavier contributions are suppressed by their large masses squared.

We reiterate that the GFFs can impose independent constraints on the chiral invariant mass.
Importantly, future physics programs at the Electron–Ion Collider, aimed at high precision measurements of GFFs, will play a crucial role in constraining the chiral invariant mass of various hadrons.
While the study of GFFs is fundamentally important from the viewpoint of elementary particle physics, these mass constraints would also be highly valuable for astrophysics, particularly NS physics.
Furthermore, the present framework is readily applicable to other hadrons: once the $D$-form factor of a given hadron is determined, its chiral invariant mass can likewise be identified. The resulting estimate of the coupling constant $g_{HH\sigma}$ then provides new constraints and insights into the physics of hadron–hadron interactions mediated by scalar meson dynamics. Overall, our finding bridges the investigation of GFFs with astrophysics and hadron physics.

Finally, we remark that the decomposition of the nucleon mass in this work is formulated at the level of hadronic degrees of freedom. In contrast, the underlying QCD describes the nucleon mass in terms of quark and gluon operators included in the QCD Lagrangian~\cite{Ji:1994av,Lorce:2017xzd}.
A conceptual gap remains between these two descriptions. Although bridging this gap---namely, clarifying the origin of the chiral invariant mass---remains an important open problem, our analysis indicates that a substantial portion of the nucleon mass originates from the chiral invariant dynamics of nonperturbative QCD.
\\

{\it Acknowledgments}---The work of M.K. is supported by RFIS-NSFC under Grant No. W2433019.
The work of M.H. is supported in part 
by Grants-in-Aid for Scientific Research under Grant Numbers 23H05439, 24K07045. Y.~L. M. is supported in part by the National Science Foundation of China (NSFC) under Grant No. 12547104 and Gusu Talent Innovation Program under Grant No. ZXL2024363.

\bibliography{reference.bib}

\end{document}